\documentclass[a4paper,10p,preprint,twocolumn,3p]{elsarticle}
\usepackage{hyperref}
\usepackage{amsmath}
\journal{Photoacoustics}
\usepackage{siunitx}

\begin{document}

\begin{frontmatter}

%
\title{Detection, numerical simulation and approximate inversion of 
optoacoustic signals generated in multi-layered PVA hydrogel based tissue 
phantoms}
\author{E.\, Blumenr\"other}
\ead{elias.blumenroether@hot.uni-hannover.de}
\author{O.\, Melchert}
\ead{oliver.melchert@hot.uni-hannover.de}
\author{M.\, Wollweber}
\author{B.\, Roth}
%
\address{Hannover Centre for Optical Technologies (HOT), 
Interdisciplinary Research Centre of the Leibniz Universit\"{a}t Hannover, 
Nienburger Str.\,17, D-30167 Hannover, Germany}
%

\begin{abstract}
In this article we characterize optoacoustic signals generated from layered 
tissue phantoms via short laser pulses by experimental and numerical means. 
In particular, we consider the case where 
scattering is effectively negligible and the absorbed energy density follows
Beer-Lambert's law, i.e.\ is characterized by an exponential decay within the
layers and discontinuities at interfaces.  We complement experiments on
samples with multiple layers, where the material properties are known 
{\emph{a priori}}, with numerical calculations for a pointlike detector, 
tailored to suit our experimental setup.
Experimentally, we characterize the acoustic signal observed by a piezoelectric
detector in the acoustic far-field in backward mode and we discuss the
implication of acoustic diffraction on our measurements.  We further attempt an
inversion of an OA signal in the far-field approximation.
\end{abstract}

\begin{keyword}
Optoacoustics \sep PVA hydrogel phantom \sep approximate signal inversion
\PACS 78.20.Pa \sep 07.05.Tp \sep 43.38.+n
\end{keyword}

\end{frontmatter}

\section{Introduction}

Recent progress in the field of optoacoustics (OAs) has been driven by
tomography and imaging applications in the context of biomedical optics.
Motivated by their immediate relevance for medical applications, high
resolution scans on living tissue proved the potential of the optical
absorption based measurement technique
\cite{Wang:2012,Stoffels:2015,Stoffels:2015ERR}.  Requiring a multitude of
detection points around the source volume, OA tomography allows for the
reconstruction of highly detailed images,
see, e.g., Ref.\,\cite{Gateau:2013}, assuming a mathematical model that
mediates the underlying diffraction transformation of OA signals
in the {\emph{forward}} direction \cite{Wang:2009,Colton:2013,Kuchment:2008,Kruger:1995}.

However, for most cases of {\emph{in vivo}} measurements, especially on humans,
it is not feasible to place ultrasound detectors in opposition to the
illumination source (with the ``object'' in between), i.e.\ to work in
{\emph{forward mode}}. Instead, it is worked in {\emph{backward mode}}, where
detector and irradiation source are positioned on the same side of the sample.
Using elaborate setups combining the paths of light and sound
waves it is possible to co-align optical and acoustic focus within the sample.
By scanning over a multitude of detection points it is then possible to produce
$3$D images with very high resolution, see, e.g., Ref.\,\cite{Zhang:2006}. 

A conceptually different approach is to perform measurements by means of a
single, unfocused transducer.  Albeit it is not possible to reconstruct OA
properties of arbitrary $3$D objects with a fixed irradiation source and a single
detection point only, useful information of the internal material properties
of, say, layered samples can be gained nevertheless.  In this regard, acoustic
near-field measurements by means of a transparent optoacoustic detector where
shown to reproduce the depth profile of absorbed energy density and absorption
coefficient without the need of extensive postprocessing
\cite{Jaeger:2005,Niederhauser:2005,Paltauf:2000,Paltauf:2002}.  However,
requiring close proximity and plane wave symmetry, near-field conditions are unrealistic considering most measurement scenarios. 
In contrast, the acoustic far-field regime allows for a much higher
experimental flexibility, although at the cost of the straight forward
interpretation of the measurements. More precisely, in the far-field, when the
distance between detector and source is large compared to the lateral extend of
the source, OA signals exhibit a diffraction-transformation which is
characteristic for the underlying system 
parameters \cite{Karabutov:1998,Sigrist:1986,Diebold:1990,Diebold:1991}.
In particular, in the acoustic far-field, OA signals exhibit a train 
of compression and rarefaction peaks and phases, signaling a
sudden change of the absorptive characteristics of the underlying layered
structure. 

In the presented article we thoroughly prepare and analyze polyvinyl
alcohol hydrogel (PVA-H) phantoms, comprised of layers doped with different concentrations of melanin. The acoustic properties of the PVA-Hs match those of soft tissue,
i.e.\ human skin \cite{Kharine:2003,Moran:1995}. Note that melanin is the main
endogenous absorber in the epidermis \cite{Smit:2011}, and, more importantly, 
in melanomas. 
Layers with higher concentrations of melanin absorb a greater amount
photothermal energy and expand more intensely than surrounding layers with 
low concentrations. The stress waves emitted by these OA sources, 
detected in the acoustic far field after experiencing a shape transformation 
due to diffraction, are put under scrutiny here.
Therefore, experimental measurements are complemented by custom numerical 
simulations. Besides analytic theory and experiment, the latter form a 
``third pillar'' of contemporary optoacoustic studies \cite{Wang:2009}.

The paper is organized as follows.  In Sec.~\ref{sec:TheoryAndNumerics} we
recap the theoretical background of optoacoustic signal generation and
detail our numerical approach to compute the respective signals for
point-detectors. In Sec.~\ref{sec:MethodsAndMaterial} we describe our 
experimental setup and elaborate on the preparation of the tissue phantoms
used for our measurements, followed by details of the experiments and
complementing simulations in in Sec.~\ref{sec:Results}.  We summarize
our findings in Sec.~\ref{sec:Summary}.

\section{Theory and numerical implementation}
\label{sec:TheoryAndNumerics}

We briefly recap the general theory of optoacoustic signal
generation in Subsec.\ \ref{subsect:PASigGen}. Subsequently, in Subsec.\ 
\ref{subsect:LayeredMedia}, we customize the general optoacoustic poisson 
integral to properly represent the layered tissue phantoms and irradiation
source profile used in our experiments. Finally, in Subsec.\ 
\ref{subsect:Numerics}, we emphasize some important implications of
the problem-inherent symmetries on our numerical implementation.  

\subsection{General optoacoustic signal generation}
\label{subsect:PASigGen}
In thermal confinement, i.e.\ considering short laser pulses with pulse
duration significantly smaller than the thermal relaxation time of the
underlying material \cite{Kruger:1995,comment:ThermalConfinement}, the
inhomogeneous optoacoustic wave equation relating the scalar pressure field
$p(\vec{r},t)$ to a heat absorption field $H(\vec{r},t)$ reads
\begin{equation}
\Big[ \partial_t^2 - c^{2} \vec{\nabla}^2 \Big]~p(\vec{r},t) = \partial_t~\Gamma H(\vec{r},t). \label{eq:WaveEq}
\end{equation}
Therein, $c$ signifies the speed of sound and $\Gamma$ refers to the
Gr\"uneisen parameter, an effective parameter summarizing various macroscopic
material constants, describing the fraction of absorbed
heat that is actually converted to acoustic pressure.  As evident from Eq.\
(\ref{eq:WaveEq}), temporal changes of the local heat absorption field serve as
sources for stress waves that form the optoacoustic signal. Following the
common framework of stress confinement \cite{Wang:2009}, we consider a product
ansatz for the heating function in the form
\begin{equation}
H(\vec{r},t)~=~W(\vec{r}) \delta(t), \label{eq:HeatingFunction}
\end{equation}
where $W(\vec{r})$ represents the volumetric energy density deposited in the
irradiated region due to photothermal heating by a laser pulse \cite{Tam:1986},
which, on the scale of typical acoustic propagation times, is assumed short
enough to be represented by a delta-function. 
Consequently, an analytic solution for the optoacoustic pressure at the field
point $\vec{r}$ can be obtained from the corresponding Greens-function in free
space, yielding the optoacoustic Poisson integral
\cite{DeanBen:2012,Burgholzer:2007,Landau:1987}
\begin{equation}
p(\vec{r},t) = \frac{\Gamma}{4\pi c} \partial_t
\int\limits_{V}\!\frac{W(\vec{r}^\prime)}{|\vec{r}-\vec{r}^\prime|}
\delta(|\vec{r}-\vec{r}^\prime| - ct)\,\mathrm{d}\vec{r}^\prime
,\label{eq:PAPoissonEq}
\end{equation} 
where $V$ denotes the ``source volume'' beyond which 
$W(\vec{r}^\prime)=0$ \cite{comment:PrincipalValue}, and $\delta(\cdot)$
limiting the integration to a time-dependent surface constraint by
$|\vec{r}-\vec{r}^\prime| = ct$.

\subsection{The Poisson integral for layered media in cylindrical coordinates}
\label{subsect:LayeredMedia}
As pointed out earlier, we consider non-scattering compounds, composed of
(possibly) multiple plane-parallel layers, stacked along the $z$-direction of
an associated coordinate system.  Whereas the acoustic properties are assumed
to be constant within the medium, the optical properties within the absorbing
medium might change from layer to layer. Thus, the volumetric energy density
naturally factors according to 
\begin{equation}
W(\vec{r})~=~f_0 f(x,y) g(z), \label{eq:W}
\end{equation}
wherein $f_0$ signifies the energy fluence of the incident laser beam on the
$z=0$ surface of the absorbing material, and $f(x,y)$ and $g(z)$ specify the
two-dimensional ($2$D) irradiation source profile and the $1$D axial absorption
depth profile, respectively.  Bearing in mind that we consider non-scattering
media, the latter follows Beer-Lambert's law, i.e.\
\begin{equation}
g(z) = \mu_{\rm a}(z) \exp\Big\{-\int_0^z\!\mu_{\rm
a}(z^{\prime})\,\mathrm{d}z^{\prime}\Big\}, \label{eq:g}
\end{equation}
wherein $\mu_{\rm a}(z)$ denotes the depth-dependent absorption coefficient.

\begin{figure}[t!]
\begin{center}
\includegraphics[width=1.0\linewidth]{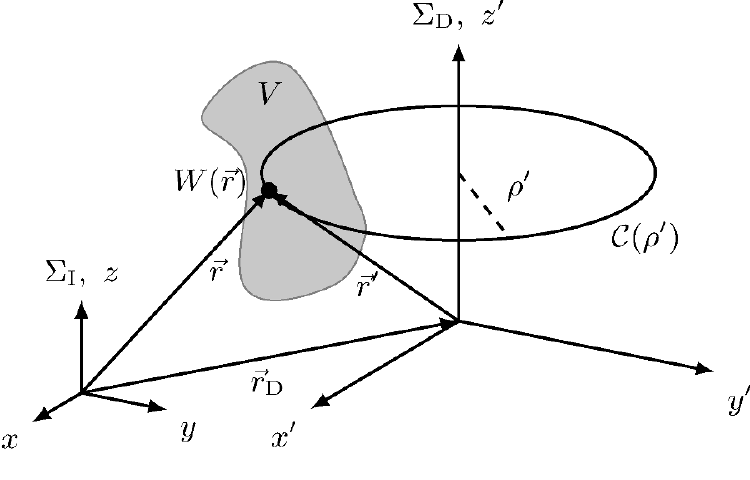}
\end{center}
\caption{
Illustration of the two reference frames $\Sigma_{\rm I}$, with origin on the
beam axis at the surface of the absorbing medium, and, $\Sigma_{\rm D}$ with
origin right at the detection point. Both coordinate systems are related by the
transformation $\vec{r}^{\prime}(\rho^{\prime},\phi^{\prime},z^{\prime})=
\vec{r} - \vec{r}_{\rm D}$ \cite{comment:Jacobian}.  Considering cylindrical
polar coordinates in $\Sigma_{\rm D}$ allows to factor the volumetric energy
density $W(\vec{r})$ within the source volume $V$ as detailed in the text and
to pre-compute the contribution of the irradiation source profile along closed
polar curves $\mathcal{C}(\rho^\prime)$ with radius $\rho^\prime$. This in turn
yields an efficient numerical scheme to compute the optoacoustic signal
$p_{\rm D}(t)$ at the detection point $\vec{r}_{\rm D}$.
\label{fig:refFrames}}
\end{figure}

Note that, for a plane-normal irradiation source with an axial symmetry, there
are two useful auxiliary reference frames based on cylindrical polar
coordinates: 
(i) $\Sigma_{\rm I}$ where $\vec{r} = \vec{r}(\rho,\phi,z)$ with origin on the
beam axis at the surface of the absorbing medium, and, (ii) $\Sigma_{\rm D}$
where $\vec{r}^\prime = \vec{r}^\prime(\rho^\prime,\phi^\prime,z^\prime)$ with
origin at the detection point $\vec{r}_{\rm D}=(x_{\rm D},0,z_{\rm D})$ in $\Sigma_{\rm
I}$, see Fig.\ \ref{fig:refFrames}. Both reference frames are related by the
point transformation $\vec{r}^{\prime}(\rho^{\prime},\phi^{\prime},z^{\prime})=
\vec{r} - \vec{r}_{\rm D}$ \cite{comment:Jacobian}.

In $\Sigma_{\rm I}$ the irradiation source profile takes the convenient form
$f(x(\rho,\phi),y(\rho,\phi))\equiv f_{\rm I}(\rho)$, where beam-profiling
measurements for our experimental setup are consistent with a top-hat shape
\begin{equation}
f_{\rm I}(\rho) = 
\begin{cases} 
  1,                             & \text{if } \rho \leq a \\ 
  \exp\{-(\rho - a)^2/d^2\},     & \text{if } \rho    > a
\end{cases}. \label{eq:f_I}
\end{equation}

In $\Sigma_{\rm D}$ the constituents of the volumetric energy density read
$f_{\rm D}(\rho^\prime,\phi^\prime) \equiv f(x_{\rm D}+\rho^\prime
\cos(\phi^\prime),\rho^\prime \sin(\phi^\prime))$ and $g_{\rm D}(z) \equiv
g(z^{\prime}-z_{\rm D})$, so that the optoacoustic Poisson integral, i.e.\
Eq.\ (\ref{eq:PAPoissonEq}), takes the form 
\begin{multline}
p_{\rm D}(t) = \frac{f_0 \Gamma}{4\pi c} \partial_t
\iiint\limits_{V}\!\rho\frac{f_{\rm D}(\rho,\phi) g_{\rm D}(z)}{(\rho^2 +
z^2)^{1/2}} \\ \times \delta((\rho^2 + z^2)^{1/2} -
ct)\,\mathrm{d}\rho\,\mathrm{d}\phi\,\mathrm{d}z. \label{eq:PAPoissonEq_D}
\end{multline}
Albeit the non-canonical formulation of the Poisson integral in cylindrical
polar coordinates might seem a bit counterintuitive at first, it paves the way
for an efficient numerical algorithm for the calculation of optoacoustic
signals for layered media. 

\subsection{Numerical experiments}
\label{subsect:Numerics}
\paragraph{Implementation details}
Considering a partitioning of the radial coordinate into $N_{\rho}$ equal sized
values $\Delta \rho = L_\rho/N_{\rho}$ so that $\rho_i=i \Delta\rho$ with
$i=0\ldots N_\rho-1$, the preceding factorization of the volumetric energy
density $W(\vec{r})$ in $\Sigma_{\rm D}$ allows to pre-compute the contribution
of the irradiation source profile in Eq.\,(\ref{eq:PAPoissonEq_D}) along closed
polar curves $\mathcal{C}(\rho_i)$ with radius $\rho_i$ according to
\begin{equation}
F_{\rm D}(\rho_i) = \lim_{N_\phi\to\infty} \rho_i \sum_{j=0}^{N_\phi-1} f_{\rm
D}(\rho_i,\phi_j)~\Delta\phi, \label{eq:F_D}
\end{equation} 
where $\Delta \phi = 2\pi/N_{\phi}$ and $\phi_j=j\Delta\phi$ with
$j=0\ldots N_{\phi}-1$, thus completing the integration over the azimuthal
angle and providing the results in a tabulated manner with time complexity
$O(N_\rho N_\phi)$.  This in turn yields an efficient numerical scheme to
compute the optoacoustic signal $p_{\rm D}(t)$ at the detection point
$\vec{r}_{\rm D}$ since the pending integrations can, in a discretized setting
with $\Delta z=L_z/N_z$ so that $z_k=k\Delta z$ for $k=0\ldots N_z-1$,
be carried out with time complexity $O(N_\rho N_z)$. Consequently, interpreting
the $\delta$-distribution in Eq.\,(\ref{eq:PAPoissonEq_D}) as an indicator
function that bins the values of the integrand according to the propagation
time of the associated stress waves, the overall algorithm completes in time
$O(N_\rho N_\phi + N_\rho N_{z})$. Note that for the special case $x_{\rm
D}=y_{\rm D}=0$, i.e.\ for detection points on the beam axis, Eq.\,(\ref{eq:F_D})
further simplifies to $F_{\rm D}(\rho_i)=2\pi \rho_i f_{\rm I}(\rho_i)$,
reducing the time complexity to only $O(N_\rho N_z)$ \cite{comment:VolterraEq}.
During our numerical simulations 
 \footnote{A Python implementation of our code for the solution of the
photoacoustic Poisson equation in cylindrical polar coordinates, i.e.\ Eq.\
(\ref{eq:PAPoissonEq_D}), can be found at \cite{Melchert_GitHub:2016}.}, for
practical purposes and since we are only interested in the general shape of the
optoacoustic signal in order to compare them to the transducer response, we set
the value of the constants in Eq.\,(\ref{eq:PAPoissonEq_D}) to $f_0 \Gamma/c
\equiv 4\pi$.  Thus, the resulting signal is obtained in
arbitrary units, subsequently abbreviated as $[{\rm a.u.}]$, making it
necessary to adjust the amplitude of the signal if we intent to compare the
results to actual measurements.  Further, to mimic the finite thickness $\Delta
w$ of the transducer foil, see Sec.\ \ref{sec:MethodsAndMaterial}, we averaged
the optoacoustic signal at the detection point over a time interval $\Delta t
= \Delta w/c$.

\begin{figure}[t!]
\begin{center}
\includegraphics[width=1.0\linewidth]{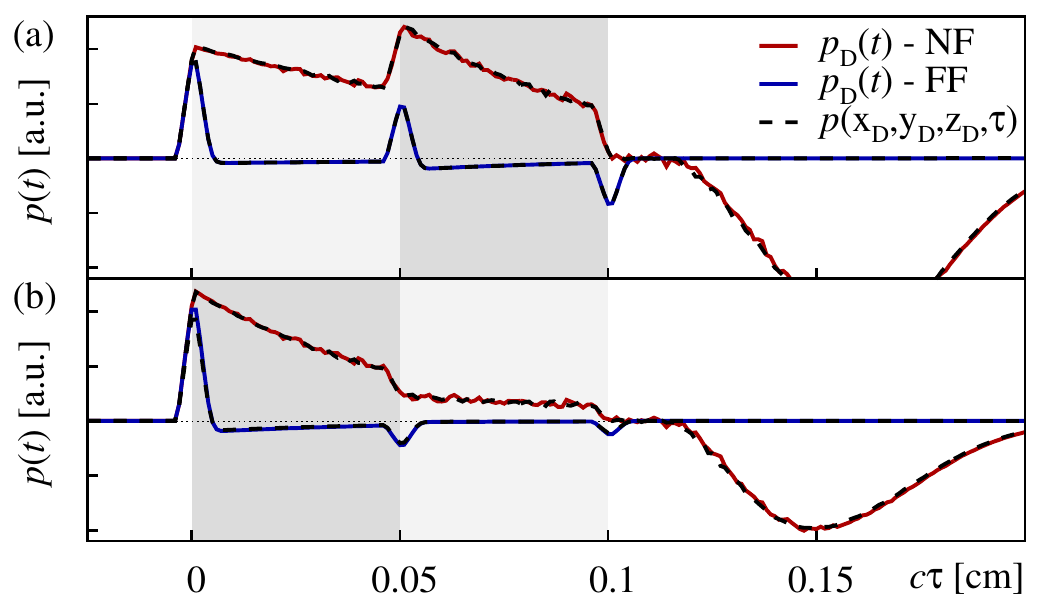}
\end{center}
\caption{
(Color online) Comparison of different solvers for the optoacoustic problem
for layered media.  The data curves labeled by $p_{\rm D}(t)$ refer to an
implementation in cylindrical polar coordinates according to Eq.\
\ref{eq:PAPoissonEq_D}. The curves are computed for a field point in the
acoustic near-field (NF; red line) and far-field (FF; blue line) at $z_{\rm
D}=-0.04~{\rm cm}$ and $z_{\rm D}=-4.0~{\rm cm}$ on the beam axis,
respectively. The corresponding numerical results obtained using an
implementation of Eq.\ \ref{eq:PAPoissonEq} in Cartesian coordinates are
labeled by $p(\vec{r}_{\rm D},\tau)$ (black dashed lines).  (a) Setup where the
source-volume encloses two absorbing layers consisting of
$\mu_a=10~\rm{cm^{-1}}$ in the range $z=0.0-0.05~{\rm cm}$ (light-gray shaded
region) followed by $\mu_a=20~\rm{cm^{-1}}$ in the range $z=0.05-0.1~{\rm cm}$
(gray shaded region), and, (b) setup where the order of the layers is reversed. 
\label{fig:cmpSolver}}
\end{figure}

\paragraph{Exemplary optoacoustic signals}
So as to facilitate intuition and to display the equivalence of the numerical
schemes implemented according to Eqs.\,(\ref{eq:PAPoissonEq}) and
(\ref{eq:PAPoissonEq_D}) in both, the acoustic near field (NF) and far field
(FF), we illustrate typical optoacoustic signals in Fig.\ \ref{fig:cmpSolver}.
Therein, the ``cartesian'' solver was based on a voxelized cubic
representation of the source volume with side-lengths
$(L_x,L_y,L_z)=(0.6,0.6,0.15)~[{\rm cm}]$ using $(N_x,N_y,N_z)=(1500,1500,150)$
meshpoints, whereas the solver based on cylindrical coordinates used a
decomposition of the computational domain into $(L_\rho,L_z)=(0.3,0.15)~{\rm
cm}$ and $(N_\rho,N_\phi,N_z)=(6000,360,150)$.  The parameters defining the
irradiation source profile were set to $a=0.15~{\rm
cm}$ and $d=a/4$. As finite thickness of the transducer foil we considered 
$\Delta w = 50~{\rm \mu m}$ in both setups.

The dimensionless diffraction parameter $D=2 |z_{\rm D}|/(\mu a_0^2)$
\cite{Karabutov:1998,Paltauf:2000} can be used to distinguish the acoustic near
field (NF) at $D<1$ and far field (FF) at $D>1$.  Here, we consider the
effective parameters $\mu=\langle \mu_{\rm a}(z) \rangle$ and $a_0=1.25 \cdot
a$ in case of multi-layered tissue phantoms. 
The simulations were performed at detection points on the beam axis, realizing
NF conditions with $D\approx 0.15$ at $z_{\rm D}=-0.04~{\rm cm}$ and FF
conditions with $D\approx 15.0$ at $z_{\rm D}=-4.0~{\rm cm}$.  As evident from
Fig.\ \ref{fig:cmpSolver}, the optoacoustic NF signals are characterized by an
extended compression phase in the range $c\tau=0.0-0.1~{\rm cm}$, originating
from the plane-wave part of the propagating stress wave and accurately tracing
the profile of the volumetric energy density along the beam axis, followed by a
pronounced diffraction valley for $c\tau > 0.11~{\rm cm}$. The particular shape
of the latter is characteristic for the top-hat irradiation source profile used
for the numeric experiments. 
In contrast to this, as can be seen from Fig.\ \ref{fig:cmpSolver}, the FF
signal features a succession of compression and rarefaction phases. Therein a
sudden increase (decrease) of the absorption coefficient is signaled by a
compression peak (rarefaction dip), cf.\ the sequence of peaks and dips at the
points $c\tau = 0,~0.05,~0.10~{\rm cm}$ in Figs.\ \ref{fig:cmpSolver}(a,b).
Further, the diffraction valley has caught up, forming rather shallow
rarefaction phases in between the peaks and dips \cite{comment:diffraction}.
Finally, note the excellent agreement of the signals obtained by the two
independent OA forward solvers.

\begin{figure}[t!]
\begin{center}
\includegraphics[width=1.0\linewidth]{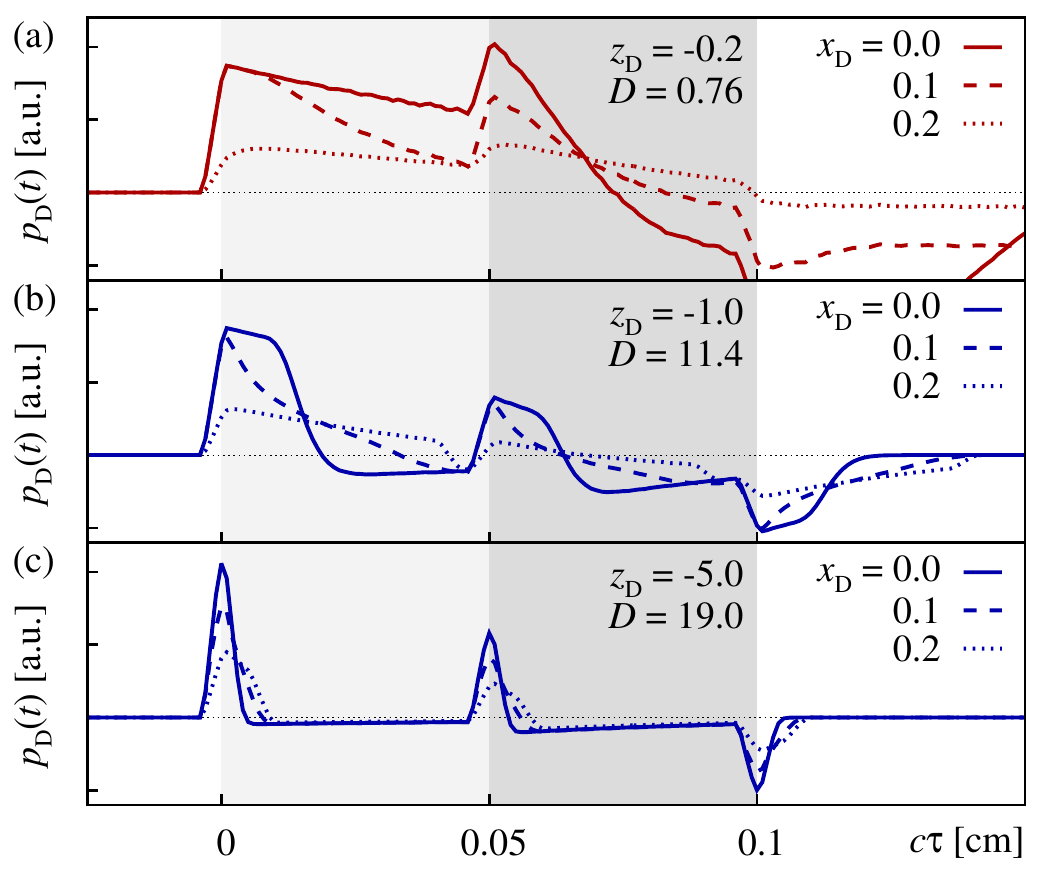}
\end{center}
\caption{
(Color online)  Sensitivity of the optoacoustic signal $p_{\rm D}(t)$ on a
radial deviation of the detection point $\vec{r}_{\rm D}$ from the beam axis,
realized by setting $x_{\rm D} \neq 0~{\rm cm}$, as explained in the text.  The
subfigures refer to different distances $z_{\rm D}$, where (a) $z_{\rm
D}=-0.2~{\rm cm}$ is located in the acoustic NF with $D=0.76$, (b) $z_{\rm
D}=-1.0~{\rm cm}$ ($D=11.4$) in the ``early'' FF, and, (c) $z_{\rm D}=-5.0~{\rm
cm}$ ($D=19.0$) in the ``deep'' FF. 
\label{fig:SONO_xDScan}}
\end{figure}

In a second series of simulations we clarified the influence of the radial
deviation of the detection point $\vec{r}_{\rm D}$ from the beam axis.
Therefore we computed the excess pressure $p_{\rm D}(t)$ at different positions
$x_{\rm D}\neq 0$ perceived in $\Sigma_{\rm I}$. The results for $x_{\rm
D}=0.1~{\rm cm}$, i.e.\ 2/3 along the flat-top part of the top-hat profile, and
$x_{\rm D}=0.2~{\rm cm}$, i.e.\ slightly above the $1/e$-width of the beam
intensity profile, are illustrated in Fig.\ \ref{fig:SONO_xDScan}.  As evident
from Fig.\ \ref{fig:SONO_xDScan}(a), for $z_{\rm D}=-0.2~{\rm cm}$, realizing a
location with $D=0.76$ in the acoustic NF, the optoacoustic signal appears to
be quite sensitive to the precise choice of $x_{\rm D}$.  I.e., as soon as the
border of the plane-wave part of the signal is approached, the transformation
of the signal due to diffraction is strongly visible.  Comparing the points
$z_{\rm D}=-1.0~{\rm cm}$ ($D=11.4$) in the ``early'' FF and $z_{\rm
D}=-5.0~{\rm cm}$ ($D=19.0$) in the ``deep'' FF, it is apparent that the
optoacoustic signal in the FF is less influenced by the off-axis deviation of
the detection point, see Figs.\ \ref{fig:SONO_xDScan}(b,c). Also, note that
with increasing distance $|z_{\rm D}|$, the interjacent rarefaction phases
level off and move closer to the leading compression peaks
\cite{comment:diffraction}.  From the above we conclude that, if we complement
actual measurements recorded in the FF via numerical simulations, we should
find a good agreement between detected and calculated signals even though both
exhibit different degrees of deviation from the beam axis.

This completes the discussion of optoacoustic signals and their generation 
from a point of view of computational theoretical physics. Details regarding
the optoacoustic detection device and the tissue phantoms are given in the
subsequent section.

\section{Methods and Material}
\label{sec:MethodsAndMaterial}
\paragraph{Photoacoustic measurement setup} 
In the following, the experimental setup is presented with focus on the phantom
preparation process and arrangement of the layered tissue samples
\cite{comment:detectorDetails}. For the detection of the OA pressure transient a self-built piezoelectric transducer is employed.
This ultrasound detector comprises a \SI{9}{\micro\meter} thick piezoelectric
polyvinylidenfluorid (PVDF) film on both sides of which $\sim$\SI{50}{nm}
indium tin oxide (ITO) electrodes are sputtered \cite{Niederhauser:2005}. The
active area of the detector is circular with a diameter of
\SI{1}{\milli\meter}. As acoustic backing layer a piece of hydrogel was
prepared and placed on top of the detector with a drop of distilled water to
ensure acoustic coupling. Due to identical acoustic impedances of the backing
layer and the phantom in addition to the marginal extent of the PVDF film in
comparison to the acoustic wavelengths, the detector can be seen as
acoustically transparent \cite{Jaeger:2005}.

\begin{figure}[t!]
\begin{center}
\includegraphics[width=1.0\linewidth]{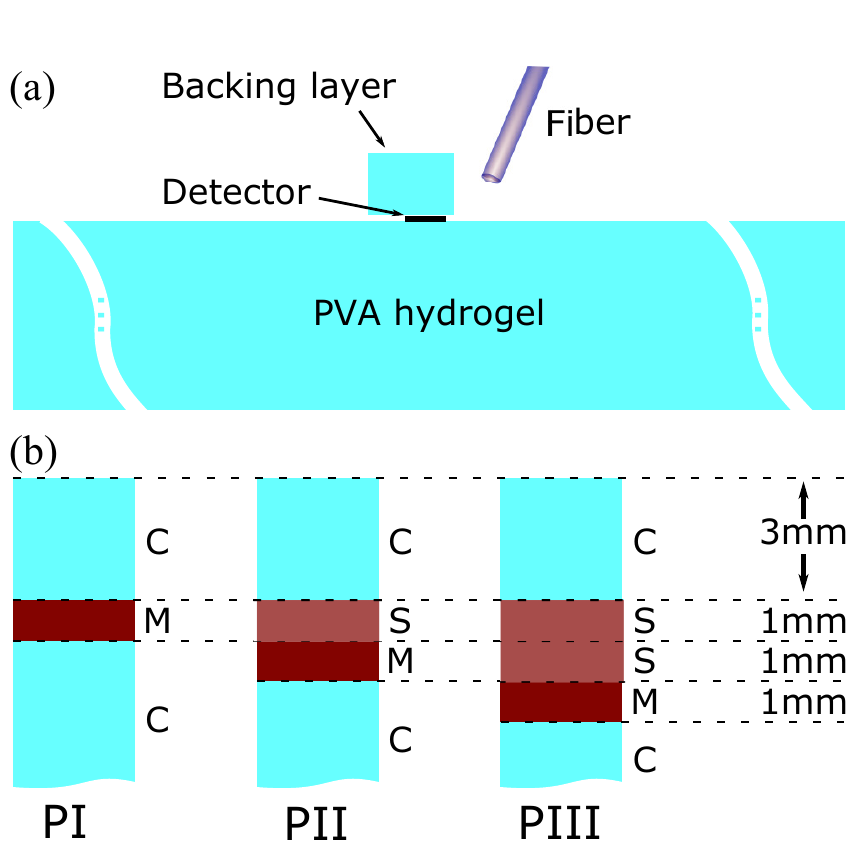}
\end{center}
\caption{
(Color online)  Sketch of the experimental setup.  (a)
Arrangement of the components as discussed in the text, see sec.\
\ref{sec:MethodsAndMaterial}. (b) Layer composition of the three different
phantoms PI, PII and PIII used for our measurements and the numerical
simulations reported in sec.\ \ref{sec:Results}.  The label ``C'' represents
clear PVA-H, ``S'' labels low absorption, and ``M'' stands for high absorption.
Note that, the clear layer at the bottom is \SI{10}{\milli\meter} thick. Thus,
signal reflections from interfaces of materials with differing acoustic
properties occur well outside the measurement range in that direction.  
\label{fig:phantoms}}
\end{figure}

In contrast to the numerical approach followed in the preceding section, the
irradiation in our experimental setup had to be adjusted to an angle
approximately \SI{20}{\degree} off the plane normal, with the light entering
the phantom in close proximity of the detector, see Fig.\ref{fig:phantoms}.  As
laser source, an optical parametric oscillator (NL303G + PG122UV, Ekspla,
Lithuania) at a wavelength of \SI{532}{\nano\meter} is coupled into a
\SI{800}{\micro\meter} fiber (Ceramoptec, Optran WF 800/880N).  The pulse
duration from the pump is 3-\SI{6}{ns}.  The beam profile measured after the 
fiber is in good agreement with a top-hat shape, which is in
accordance with the irradiation source profile Eq.\ (\ref{eq:f_I}), considered
for the previous numerical experiments, 
and parameters as detailed for the numerical simulations in Sec.\
\ref{sec:Results}.

To improve the signal-to-noise ratio and match the electrical impedance, a
custom build electrical pre-amplifier is connected to the detector electrodes.
The voltages, corresponding to the detected pressure, are recorded at 2 GSPS
(Giga sample per second) by a high-speed data acquisition card (Agilent U1065A,
up to 8GSPS). At such sampling rates, the expected ultrasound pressure profile
is highly over-sampled, thus, the point to point noise can be smoothed out
without loss of information. A conservative estimation of the fastest changing
signal features yield a time window of \SI{20}{ns} over which smoothing might be
carried out, corresponding to 40 consecutive data points.

\paragraph{Polyvinyl alcohol based hydrogel tissue phantom recipe}
The tissue phantoms used in our studies are compounds composed of stacked
layers of polyvinyl alcohol hydrogel (PVA-H)\cite{Wollweber:2014}. The
incentive to utilize PVA-H is its acoustic similarity to soft tissue, i.e.\
human skin \cite{Kharine:2003}. In contrast to liquid phantoms such as water
ink solutions \cite{Paltauf:2000}, hydrogels have the advantage of being
stackable without the need of containing walls. Furthermore, while liquids
would intermix at interfaces and thus require solid boundaries in between,
hydrogels allow sharp junctions only softened by diffusion. In the remainder
the phantom creation recipe is detailed.

Here, PVA-Hs are produced by mixing polyvinyl alcohol granulate (Sigma-Aldrich
363146, Mw 85-124 99+\% hydrolyzed) with distilled water at a mixing ratio 1:5.
Using a magnetic stirrer with heating, the dispersion is kept at
\SI{94}{\degreeCelsius} for at least \SI{40}{minutes} while the stirring bar rotates at
\SI{350}{RPM}, until it becomes a homogeneous solution.
This viscous mass can be poured into any mold to obtain the desired form.
Depending on the required thickness, a commercial metal spacing washer or a 3D
printed plastic ring of specific height was placed in between two glass plates,
thus creating very flat PVA-H cylinders.  Due to the much larger lateral extend
of the phantoms, compared to the depth of the absorbing layers, boundary
effects do not interfere with the optoacoustic signal.\\ To facilitate
polymerization the phantom is subjected to one \emph{freezing and thawing cycle}.  The
phantom is placed in the freezer at \SI{-14}{\degreeCelsius} for \SI{2}{days}.
Thawing is achieved by keeping the samples at room temperature for a few
minutes, afterwards the phantoms are ready to use. Crystallites produced by
freezing of water in the hydrogels would yield turbidity \cite{Hassan:2000}.
As a remedy, so as to obtain clear PVA-H, water soluble anti freezing agents 
are added. 
Here, when the PVA is completely dissolved, $\sim$\SI{45}{vol\percent} pure
ethanol was added to the aqueous solution incrementally, each time waiting for
the schlieren to dissolve. Especially after adding the ethanol it is very
important to keep the vessel closed whenever possible.

The optical properties of the samples can be manipulated by inclusion of
scatterers and or absorbers. In our studies, synthetic melanin (Sigma-Aldrich,
M0418-100MG) was chosen as absorber to mimic melanoma, that is, black skin
cancer. Due to its robustness to temperature, the finely ground melanin can be
included in the beginning of the phantom creation process, at the same time
with the PVA pellets. As a rough estimate it is assumed that melanomas contain
as much melanin as African skin. According to \cite{Karsten:2012}, dark skin
contains 10 times the melanin as compared to very fair skin, not taking into
account the type of melanin. So as to reproduce the contrast of a melanoma in
Caucasian skin the following different types of PVA-H layers were created:
\begin{itemize}
 \item[(i)] PVA-H without melanin, referred to as ``C'',
 \item[(ii)] PVA-H with \SI[per-mode=symbol]{1}{\milli\gram\per\milli\liter} of melanin, referred to as ``M'', and, 
 \item[(iii)] PVA-H with \SI[per-mode=symbol]{0.1}{\milli\gram\per\milli\liter} of melanin, referred to as ``S'',
\end{itemize}
By stacking these in different order, three distinct phantoms were created, see
Fig.\ \ref{fig:phantoms}. Note that, the melanin concentrations specified
above relate to the amount of hydrogel before the addition of ethanol. In the
presented study we considered non-scattering material only, thus we did not add
any scattering supplements. 

\paragraph{Further hints for handling the tissue phantoms}
Below we list practical hints and findings which proved very
helpful in the course of tissue phantom production:
\begin{itemize}
\item[H1:] 
In a closed screw neck bottle the aqueous solution can be stored for weeks at
room temperature. 
However, note that after the flask has been reheated and opened several times, 
inevitable ethanol evaporation might cause turbidity upon freezing.
\item[H2:] 
The undesirable formation of bubbles within the phantoms can be avoided, to 
a great extend, by overfilling the spacing ring and mounting the top glass 
plate on the sample after the hydrogel has settled for a while. 
This procedure prohibits trapping of air at the spacing ring boundaries as 
well as giving the hydrogel some time to degas. 
\item[H3:] 
While stacking the PVA-H layers in preparation for a measurement, the 
individual phantom layers should be kept wet by means of distilled water
in order to prevent them from sicking together with one another and, most of 
all, themselves. Also, a proper watery film prohibits the inclusion of air 
in between layers.
\end{itemize}


\section{Results}
\label{sec:Results}

\begin{figure}[th!]
\begin{center}
\includegraphics[width=1.0\linewidth]{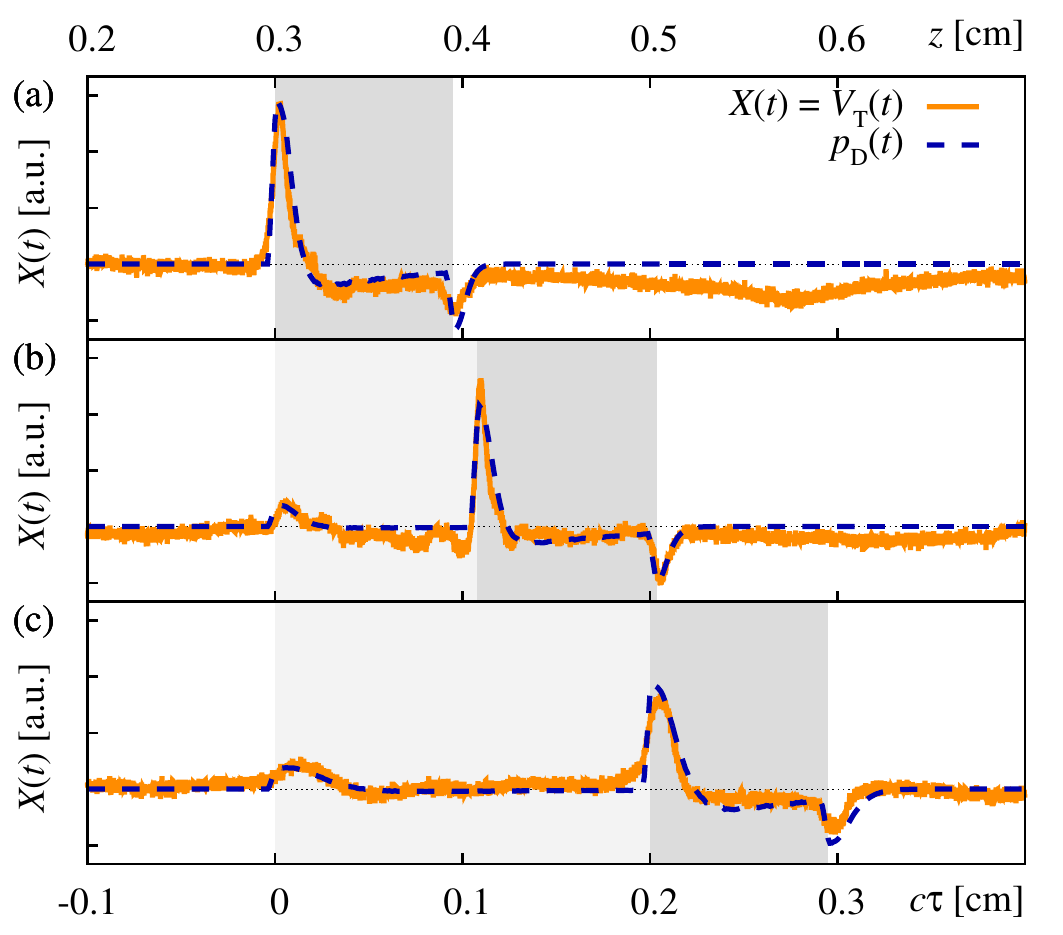}
\end{center}
\caption{
(Color online) Comparison of optoacoustic signals for layered media obtained
from measurements ( labeled ``$V_{\rm
D}(t)$''; orange solid lines) and numerical simulations (labeled ``$p_{\rm
D}(t)$''; blue dashed lines).  
The top and bottom abscissas refer to the distance $z$ traveled by the signal
and the retarded signal depth $c \tau = c t - z_{\rm D}$ (where $z_{\rm D}=0.3~{\rm cm}$),
respectively.  
(a) single-layer tissue phantom ${\rm PI}$,
(b) double-layer tissue phantom ${\rm PII}$, and, 
(c) double-layer tissue phantom ${\rm PIII}$, see Fig.\ \ref{fig:phantoms}(b).
The layer with different melanin concentrations are indicated by the light-gray
(in case of ``S''; cf.\ \ref{fig:phantoms}(b)) and gray (in case of ``M''; cf.\
\ref{fig:phantoms}(b)) shaded regions. Note, the presented data was obtained in single measurement and was neither smoothed nor averaged.
\label{fig:TvsE}}
\end{figure}  

Below we complement measured PA signals, obtained from measurements on the
three tissue phantoms ${\rm PI-III}$, discussed in Sec.\
\ref{sec:MethodsAndMaterial} and illustrated in Fig.\ \ref{fig:phantoms}(b),   
with custom simulations obtained in terms of the numerical framework detailed
in Sec.\ \ref{sec:TheoryAndNumerics}. 
As evident from the comparison of the experimental setup with the simulation
framework, there are three distinctions between experiment and theory which
have to be kept in mind while interpreting the results: 
(i) while the irradiation source is assumed to be plain normal incident for our
simulations, the direction of incidence in the experimental setup exhibits a nonzero angle off the plane normal. Additionally, due to
unavoidable refraction at the phantom surface, the beam profile is likely to be
non-symmetric and slightly divergent. Hence, the top-hat beam shape assumed in
our simulation approach can only been seen as an approximation of the
experimental conditions. 
(ii) although it is probable that all the measurements are performed,
at least to some extend, off-axis we opt for modeling and numerical simulations
in an on-axis approach. As demonstrated in Subsec.\ \ref{subsect:Numerics} and
illustrated in Fig.\ \ref{fig:SONO_xDScan}(c), we expect the principal signal
shape in the acoustic far-field to change only at a small rate upon deviation
from the beam axis.  
(iii) as pointed out in sec.\ \ref{sec:MethodsAndMaterial}, the active area of the transducer has a radius of $0.5~{\rm mm}$, while in our simulations we compute optoacoustic
signals for a pointlike detector. However, upon approaching the far-field limit
one expects the former intrinsic length scale not to be of significance. Albeit
we plan to address this issue in future work, the apparent qualitative
agreement of simulation and experiment detailed in the remainder is impressive
and should suffice to validate our approach.
\paragraph{Comparison of optoacoustic signals obtained in theory and experiment}
The measured optoacoustic signals for the tissue phantoms ${\rm PI-III}$ along
with the simulated curves are illustrated in Figs.\ \ref{fig:TvsE}(a-c).  In
principle all three measured curves exhibit the characteristic features expected for signals observed in the acoustic far-field. Thus
these measurements are well suited for the purpose of optoacoustic depth
profiling \cite{Paltauf:2000}.
In particular, for the simulation of ${\rm PI}$ we considered a single layer
with absorption coefficient $\mu_a=11~{\rm cm^{-1}}$ in the range
$z=0.3-0.395~{\rm cm}$ (note that $z$ is measured with respect to the origin of
$\Sigma_{\rm D}$), indicated by a gray shaded region representing a highly
absorbing layer (introduced as ``M'' in sect.\ \ref{sec:MethodsAndMaterial}).
The top-hat beam shape parameters within the simulation where set to
$a=0.054~{\rm cm}$ and $R\equiv d/a = 1.5$.  Note that both principal features
of the signal, i.e.\ the initial compression peak as well as the trailing
rarefaction dip are reproduced well and the intermediate rarefaction phase
matches well in theory and experiment.  The subsequent long and shallow
rarefaction phase for $z>0.5~{\rm cm}$ is located outside the measurement range
that corresponds to the prepared source volume and is likely caused by acoustic
reflections from the lateral boundaries of the backing layer.

The same holds for the analysis of the remaining two phantoms, where ${\rm
PII}$ was modeled by considering a first layer with a comparatively low
absorption coefficient $\mu_a=1.4~{\rm cm^{-1}}$, i.e.\ type-S, in the range
$z=0.3-0.408~{\rm cm}$ (light-gray shaded region), followed by a type-M layer
with $\mu_a=11~{\rm cm^{-1}}$ in the range $z=0.408-0.504~{\rm cm}$ (gray
shaded region). Therein, the beam shape parameters where set to $a=0.056~{\rm
cm}$ and $R=1.2$. Here, all three expected characteristic signal
features, i.e.\ the initial small compression peak, the interjacent high
compression peak as well as the trailing rarefaction dip match well for theory
and experiment.

Finally, phantom ${\rm PIII}$ was modeled by considering a type-S layer with
$\mu_a=1.4~{\rm cm^{-1}}$ in the range $z=0.3-0.5~{\rm cm}$ (light-gray shaded
region) followed by a type-M layer with $\mu_a=11~{\rm cm^{-1}}$ in the range
$z=0.5-0.595~{\rm cm}$ (gray shaded region). Therein, the beam shape parameters
where set to $a=0.08~{\rm cm}$ and $R=1.2$.  Again, all three
characteristic signal features are reproduced well by theory and experiment.

Note that, as pointed out in subsec.\ \ref{subsect:Numerics}, it is necessary 
to adjust the scale of the amplitude of the computed PA signal if we intend to 
compare it to the transducer response. The respective scaling factor was 
obtained from the simulated and measured curves for tissue phantom ${\rm PI}$
and was subsequently used in the other two cases to achieve the excellent agreement 
displayed in Figs.\ \ref{fig:TvsE}(a-c).

\begin{figure}[th!]
\begin{center}
\includegraphics[width=1.0\linewidth]{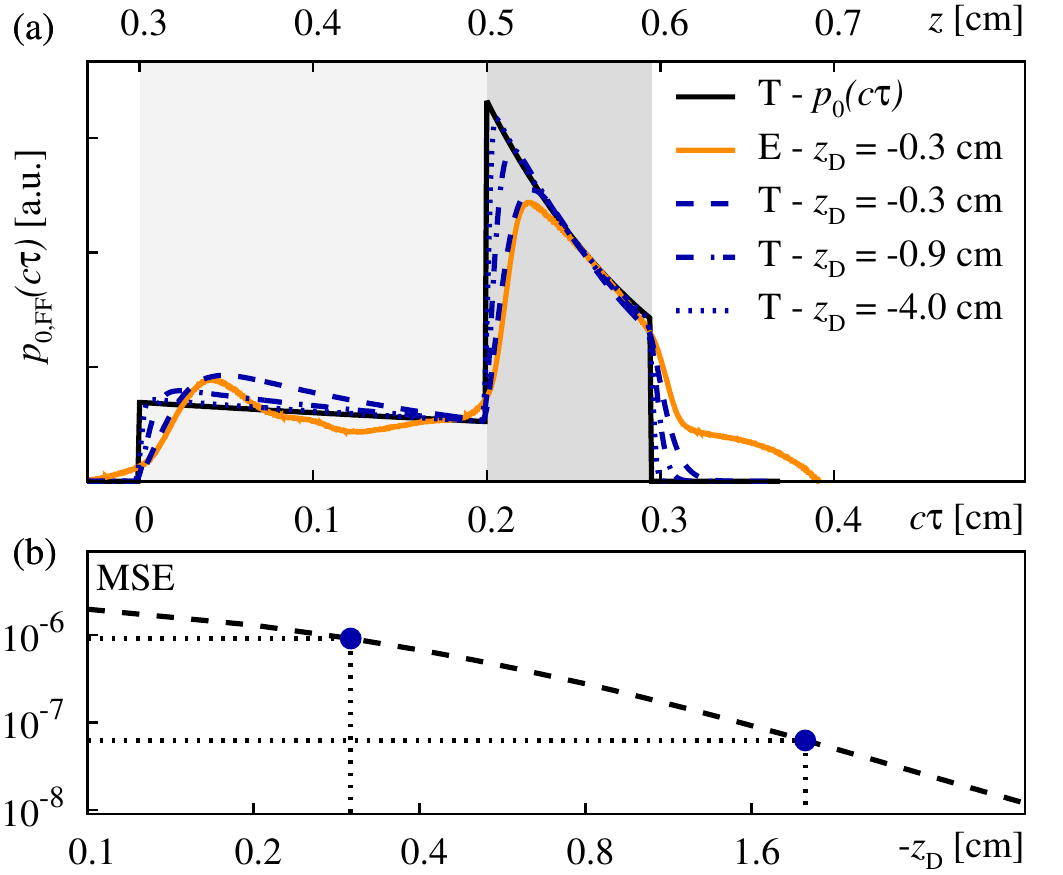}
\end{center}
\caption{
(Color online) Reconstruction of the initial volumetric energy density in the 
far-field (FF) approximation for tissue phantom ${\rm PIII}$, see Fig.\ 
\ref{fig:TvsE}(c).
(a) comparison of the exact initial distribution of acoustic stress $p_{\rm 0}$ 
(black solid line; labeled ``T'') to 
FF reconstructed predictors $p_{\rm 0,FF}$ simulated at different 
measurement points $z_{\rm D}$ (blue lines; labeled ``T'') 
and the FF reconstructed predictor derived from our measurement (orange solid 
line; labeled ``E''). 
(b) mean square error ${\rm MSE}$ between the exact initial volumetric 
energy density and the FF reconstructed value from the optoacoustic 
signals calculated at different detection points at $z_{\rm D}=-0.1$ 
through $-4$ ${\rm cm}$.
\label{fig:p0Predictor_PIII}}
\end{figure}  

\paragraph{Reconstruction of the initial volumetric energy distribution}

As discussed in the literature, in the acoustic far-field, the observed 
optoacoustic signal can be related to the initial volumetric energy density
by means of a temporal derivative \cite{Paltauf:2000,Karabutov:1998}.
Consequently, as discussed in Ref.\ \cite{Paltauf:2000}, this offers the 
possibility to reconstruct the initial acoustic stress distribution 
$p_{\rm 0}(z) = \Gamma\, W(z)$ in the limit $D\gg 1$. Albeit Ref.\
\cite{Paltauf:2000,Paltauf:2002} used the integral of the measured acoustic
signals as a visual aid for imaging purposes, cf.\ Fig.\ 9(c) of Ref.\
\cite{Paltauf:2000}, and Fig.\ 8(b) of Ref.\ \cite{Paltauf:2002}, they did not
elaborate on this issue any further.
Albeit we agree that the FF signals are naturally suited for the purpose of
optoacoustic depth profiling, we here attempt to explore the use of the above
idea in order to obtain a predictor $p_{\rm 0,FF}\approx p_{\rm 0}$ in terms of
a FF approximation for tissue phantom ${\rm PIII}$.
This is illustrated in Fig.\ \ref{fig:p0Predictor_PIII}(a), where we 
show the exact initial distribution of acoustic stress $p_{\rm 0}$ (solid black line) by 
means of which the numerical simulations where carried out, together with 
the FF reconstructed predictors $p_{\rm 0,FF}$ simulated at three different 
measurement points $z_{\rm D}=-0.3,~-0.9,~-4.0~{\rm cm}$ in the acoustic FF
and the FF reconstructed predictor derived from our measurement. 
While the measurement based and simulation based predictors at 
$z_{\rm D}=-0.3~{\rm cm}$ agree quite well it can be seen that, even though
the simulations are carried out in acoustic FF, they still differ noticeably
from the exact curve. As one might intuitively expect, an increasing distance
$|z_{\rm D}|$ yields a more consistent estimate. In the limit $|z_{\rm D}|\to \infty$
this is limited only by the temporal averaging of the signal, implemented to mimic 
a finite thickness of the transducer foil.

This can be assessed on a quantitative basis by monitoring the mean squared
error ${\rm MSE} = \sum_{i=0}^{N_z-1} [ p_{\rm 0}(z_i) - p_{\rm 0,FF}(z_i)]^2/N_z$ in 
a discretized setting, with $z_i$ as in Subsec.\ 
\ref{subsect:Numerics}, see Fig.\ \ref{fig:p0Predictor_PIII}(b).
Note that in advance, the above signals are normalized in order to ensure
$\sum_i X(z_i)=1$ for both, $X=p_{\rm 0}$ and $p_{\rm 0,FF}$. 
As evident from the figure, the MSE might be reduced by a solid order of magnitude
upon moving the signal detection from $z_{\rm D}=-0.3~{\rm cm}$ to $-2.0~{\rm cm}$
further into the far-field (indicated by the dashed lines in the figure). 

\section{Summary and Conclusions}
\label{sec:Summary}

In the presented article we discussed an efficient numerical procedure for the
calculation of optoacoustic signals in layered media, based on a numerical
integration of the optoacoustic Poisson integral in cylindrical polar
coordinates, in combination with experimental measurements on PVA based
hydrogel tissue phantoms.
In summary, we observed that far-field measurements on tissue phantoms composed
of layers with different concentrations of melanin are in striking agreement
with custom numerical simulations and exhibit all the characteristic features
that allow for optoacoustic depth profiling. Further, in our experiments, the
signal to noise ratio of {\emph single} measurements was sufficiently high to
omit any signal post-processing. 
In contrast to the experimental measurements, the simulations are performed
with on axis illumination and assuming an ideal pointlike detector.
Nonetheless, simulation and experiment agree very well over all, which
highlights the robustness of the signal analysis and simulation against small
deviations.
Finally, we showcased the possibility to reconstruct the initial pressure
profile in a far-field approximation by numerical integration. Even though
exact reconstruction would require an ideal detector in addition to an infinite
distance between source and detector, the pressure profile reconstructed here
(at finite distance $|z_{\rm D}|=1~{\rm cm}$ and finite detector radius
$0.5~{\rm mm}$) reproduces the initial pressure profile exceedingly knorke. 
In this regard, from the point of view of computational theoretical physics, it
is also tempting to explore further, conceptually different signal inversion
approaches, that might facilitate a reconstruction of ``internal''
optoacoustic material properties based on the measurement of ``external'' OA
signals. Such investigations are currently in progress.


\section*{Acknowledgments}
We thank J. Stritzel for valuable discussions and comments, as well as for
critically reading the manuscript.  We further thank M.\ Wilke for assisting in
the preparation of the PVA-H tissue phantoms.
E.\ B.\ acknowledges support from the German Federal Ministry of Education and 
Research (BMBF) in the framework of the project MeDiOO (Grant FKZ 03V0826).
O.\ M.\ acknowledges support from the 
VolkswagenStiftung within the ``Nieders\"achsisches Vorab'' program in the framework of
the project ``Hybrid Numerical Optics''  (Grant ZN 3061). 
Further valuable discussions within the collaboration of projects MeDiOO and
HYMNOS at HOT are gratefully acknowledged.

\bibliography{masterBibfile_optoacoustics,commentsBibfile_optoacoustics}

\begin{thebibliography}{10}
\bibitem{comment:PrincipalValue}
Note that, so as to exclude the nonphysical effect of a point source on itself,
  the Poisson integral has to be understood in terms of its principal value if
  the source volume $V$ contains the field point $\vec{r}$.
\bibitem{comment:Jacobian}
Note that the Jacobian determinant that mediates the transformation from the
  cartesian $\Sigma_{\rm I}$ to the cylindrical $\Sigma_{\rm D}$ merely reads
  $\det(J(x,y,z;\rho^\prime,\theta^\prime,z^\prime))=\rho^{\prime}$.
\bibitem{comment:diffraction}
Note that the diffraction stress-wave appears more shallow in the FF since the
  photoacoustic Poisson integral, i.e.\ Eq.\ (\ref{eq:PAPoissonEq}), implies
  that the amplitude decreases inversely proportional to the propagation
  distance of the wave.
\bibitem{comment:detectorDetails}
An in depth description of the detector design and layout is out of scope of
  the presented study and will be discussed elsewhere.
\end{thebibliography}

\end{document}